\documentclass[conference]{IEEEtran}

\usepackage{graphicx}
\usepackage{epstopdf}
\usepackage{amssymb}
\usepackage[cmex10]{amsmath}
\usepackage{amsmath}
\usepackage[font=footnotesize]{subfig}
\usepackage{mathtools}
\usepackage{tikz}
\usetikzlibrary{shapes}
\usepackage{pgfplots}
\tikzset{>=latex}

\usepackage{multicol}

\newtheorem{lemma}{Lemma}

\begin{document}
\bstctlcite{IEEEexample:BSTcontrol}

\title{Age and Value of Information: Non-linear Age Case}

\author{
	\IEEEauthorblockN{Antzela~Kosta\IEEEauthorrefmark{1}, Nikolaos~Pappas\IEEEauthorrefmark{1}, Anthony~Ephremides\IEEEauthorrefmark{1}\IEEEauthorrefmark{2}, and~Vangelis~Angelakis\IEEEauthorrefmark{1}}
	\IEEEauthorblockA{\IEEEauthorrefmark{1} Department of Science and Technology, Link{\"o}ping University, Campus  Norrk{\"o}ping, 
		60 174, Sweden}
	\IEEEauthorblockA{\IEEEauthorrefmark{2} Electrical and Computer Engineering Department, University of Maryland, College Park, MD 20742\\
		E-mail: \{antzela.kosta, nikolaos.pappas, vangelis.angelakis\}@liu.se,  etony@umd.edu}  }
\maketitle

\begin{abstract}
We consider a real-time status update system consisting of a source-destination network.
A stochastic process is observed at the source, and samples, so called status updates, are extracted at random time instances, and delivered to the destination.
In this paper, we expand the concept of information ageing by introducing the \emph{Cost of Update Delay} (CoUD) metric to characterize the cost of having stale information at the destination.
We introduce the \emph{Value of Information of Update} (VoIU) metric that captures the reduction of CoUD upon reception of an update.
The importance of the VoIU metric lies on its tractability which enables the minimization of the average CoUD.
\end{abstract}
\IEEEpeerreviewmaketitle
\section{Introduction}
Our work is motivated by the need for adaptability to meet stringent timeliness constraints in communication systems, arising from sensing and actuation applications within the IoT.
Characterization of  time-critical information can be done through the so called \emph{real-time status updates} that are messages carrying the timestamp of their generation.
Status updates can range from sensor observations to stock market data, tracking time-varying content over a network. 

A common objective of such communications systems is to maximize the freshness of the received data.
We consider a stochastic process being observed by a source that extracts samples or \emph{status updates} at random times. 
The status updates are transmitted over a network in order to update the destination node about the evolution of the process.

To quantify freshness, the concept of \emph{Age of Information} (AoI) has been introduced in \cite{Kaul12_INFOCOM}, to characterize the timeliness of information in a status update system.
The \emph{age} captures the elapsed time since the last received status update was generated.
More specifically, at time of observation age is defined as the current time excluding the time at which the observed state was generated.
Keeping average AoI small corresponds to maintaining fresh information.

Part of AoI research has so far focused on the use of different queuing models through which the status updates may be processed.
The average age has been investigated in \cite{Kaul12_INFOCOM} for the M/M/1, D/M/1 and M/D/1 queues.
In \cite{Kam16} the authors  take into consideration a more dynamic feature of wireless networks, that is, packets traveling over the network might reach the destination through multiple paths.
This gives rise to out-of-order delivery and thus non-informative (obsolete) packets \cite{Kam13_ISIT}.
The performance of the M/M/1, M/M/2, and M/M/$\infty$ cases is investigated in \cite{Kam16}.
Multiple sources are studied in \cite{Yates12_ISIT}, \cite{Yates17}, where the authors characterize how the service facility can be shared among multiple update sources. 
In \cite{Costa14_ISIT}, the new metric of \emph{peak age of information} (PAoI) was introduced.
In \cite{Modiano15_ISIT}, the authors consider the problem of optimizing the PAoI by controlling the arrival rate of update messages and derive properties of the optimal solution for the M/G/1 and M/G/1/1 models.

Controlling the messages in a network can increase the performance, starting from a simple last-generated-first-served (LGFS) service discipline \cite{Kaul12_CISS}, to more complicated packet management that discards non informative packets \cite{Costa14_ISIT}, \cite{Costa16}, \cite{Pappas15_ICC}.
In \cite{Kam16_ISIT} the authors introduce packet deadlines as a control mechanism and study its impact on the average age of a M/M/1/2 system.
In \cite{Chen16_ISIT} on the other hand, the authors take into consideration packet delivery errors, i.e., update packets can get lost during transmissions to their destination.
In \cite{Bedewy16_ISIT} the authors consider multiple servers where each server can be viewed as a wireless link.
They prove that a preemptive LGFS service simultaneous optimizes the age, throughput, and delay performance in infinite buffer queuing systems.
In \cite{Bedewy17arxiv} the minimization of age is done over general multihop networks.
Another control policy is to assume that the source is monitoring the network server’s idle/busy state and is able to generate status updates at any time, as in \cite{Sun16_INFOCOM}, \cite{Yates15_ISIT}.

In \cite{Sun16_INFOCOM} the authors define an age penalty function of a general form, to characterize the level of \lq\lq{}dissatisfaction\rq\rq{} for data staleness.
Pushing forward, we investigate the \emph{Cost of Update Delay} (CoUD) metric for three sample case functions that can be easily tuned through a parameter.
For each case, we derive tractable expressions of the average cost for a M/M/1 model with a first-come-first-served (FCFS) queue discipline.

Although in \cite{Sun16_INFOCOM} penalty functions are said to be determined by the application, we go further and associate the cost of staleness with the statistics of the source.
Prior to defining this association, we first need to elaborate on the requirement of small AoI.
Why are we interested in small AoI?
Consider that we are observing a system at time instant $t$.
However, the most recent value of the observed process available is the one that had arrived at $t-\Delta$, for some random $\Delta$.
Now assume that the destination node wants to estimate the information at time $t$.
If the samples at $t$ and $t-\Delta$ are independent, the knowledge of $t-\Delta$ is not useful for the prediction and age simply indicates delay.
However, if the samples at $t$ and $t-\Delta$ are correlated, then the value of $\Delta$ will affect the accuracy of the prediction.
A smaller $\Delta$ can lead to a more accurate prediction.
Our work is a first step towards exploring this potential usage of AoI.

Next, we introduce a novel metric called \emph{Value of Information of Update} (VoIU) to capture the degree of importance of the information received at the destination.
A newly received update reduces the uncertainty of the destination about the current value of the observed stochastic process, and VoIU captures that reduction that is directly related to the time elapsed since the last update reception.
Following this approach, we take into consideration not only the probability of a reception event, but also the impact of the event on our knowledge of the evolution of the process.

Small CoUD corresponds to timely information while VoIU represents the impact of the received information in reducing the CoUD.
Therefore, in a communication system it would be highly desirable to minimize the average CoUD, and at the same time maximize the average VoIU.
To this end we obtain the average VoIU for the M/M/1 queue and discuss how the optimal server utilization with respect to VoIU can be used combined with the CoUD average analysis.
\section{System Model}
\label{sec:model}
We consider a system in which a source generates status updates in the form of packets, transmitted through a network to a remote destination.
The generated packets are queued before transmitted over the network and are transmitted according to a FCFS queue discipline.

Upon reception of a new status update, Yates et al. \cite{Kaul12_INFOCOM} defined \emph{Age of Information} $\Delta(t) = t - u(t)$ to be the difference of the current time instant and the time stamp of the received update. 
In this paper we expand the notion by defining \emph{Cost of Update Delay} (CoUD)
\begin{equation}
C(t) = f_s(t - u(t))
\label{eq:cost_t}
\end{equation} 
to be a stochastic process that increases as a function of time between updates \cite{Sun16_INFOCOM}. 
We introduce a non-negative, monotonically increasing category of functions $f_s(t)$, to represent the evolution of cost of update delay according to the data characteristics of the source of the new information.
This staleness metric increases with time and ensures that the status at the destination is as timely as possible based on the autocorrelation structure of the source signal.

Update $i$ is generated at time $t_i$ and is received by the destination at time $t'_i$.
The cost of information absence at the destination increases as a function $f_s(t)$ of time.
Note that age as coined by Yates is a special cost case, where the cost is counted in time units, as shown in Fig.~\ref{fig:linear_value}. 
In this paper we consider that the cost can take any form of a ``payment'' function that can also assign to it any relevant unit.

The $i$th interarrival time $Y_i = t_i - t_{i-1}$ is the time elapsed between the generation of update $i$  and the previous update generation and is a random variable.
Moreover,  $T_i = t'_i  - t_i$ is the system time of update $i$ corresponding to the sum of the waiting time at the queue and the network delay.
Note that the random variables $Y_i$ and $T_i$ are real system time measures and are independent of the way we choose to calculate the cost of update delay i.e., of $f_s(t)$.

At time $t'_i$, the cost $C(t'_i)$ is reset to $f_s(t'_i  - t_i)$ and we introduce the \emph{Value of Information of Update} (VoIU) $i$ as
\begin{equation}
V_i = \frac{f_s(t'_i -t_{i-1}) - f_s(t'_i - t_i)}{f_s(t'_i - t_{i-1})}
\label{eq:V_i}
\end{equation} 
to measure the degree of importance of the information received at the destination.
Intuitively, this metric depends on two system parameters at time of observation: (i) the cost of update delay at the destination (ii) the time that the received update was generated. 
This can be easily shown to be similarly expressed as a dependence on: (i) the interarrival time of the last two packets received (ii) the current reception time. 

The value of information is a bounded fraction that takes values in the real interval $\left[0,1\right] $, with 0 representing the minimum benefit of an update and 1 the maximum.
\begin{lemma}
In a system where status updates are instantaneously available from the source to the destination, VoIU is given by:
\begin{equation}
V_i = \lim_{t'_i \to t_i} V_i = 1.
\label{eq:property}
\end{equation} 
\end{lemma}
The interpretation of this property is that in the extreme case when the system time is insignificant and a packet reaches the destination as soon as it is generated, we assign to the VoIU metric the maximum value reflecting that the reception occurs without value loss.
\begin{figure*}[ht]
	\centering
	\subfloat[][]{
		  \scalebox{.65}{\begin{tikzpicture}[scale=0.90]
\draw[->] (0,0) -- (8.2,0) node[anchor=north] {time};
\draw[->] (0,0) -- (0,3.5) node[anchor=east] {$C(t)$};

\draw	(-0.3,0.25) node[anchor=south] {$C_0$};

 \draw[fill=gray!10] (1.4,0) -- (4.4,3) -- (4.4,2.4)-- (2,0);

\draw	(-0.5,0) node[anchor=north] {$t_0$}
           (0.2,0) node[anchor=north] {$t_1$}
		    (1.4,0) node[anchor=north] {$t_2$}
		    (2,0) node[anchor=north] {$t_3$}
		    (3.8,0) node[anchor=north] {$t_4$}
		    (6,0) node[anchor=north] {$t_{n-1}$}
		    (6.8,0) node[anchor=north] {$t_n$};
		    
\draw[->,>=stealth]    (1,0) -- (1,-0.4) node[anchor=south,below] {$t'_1$};
\draw[->,>=stealth]  (2.5,0) -- (2.5,-0.4) node[anchor=south,below] {$t'_2$};
\draw[->,>=stealth]  (4.4,0) -- (4.4,-0.4) node[anchor=south,below] {$t'_3$};
\draw[->,>=stealth]   (7.4,0) -- (7.4,-0.4) node[anchor=south,below] {$t'_n$};
		    	    		 
\draw	(0.55,0.7) node{{\scriptsize $Q_1$}}
		    (1.6,0.75) node{{\scriptsize $Q_2$}};
\draw   (3.7,0.75) node{{\scriptsize $Q_4$}}
           (7,0.7) node{{\scriptsize $Q_n$}};
           
\draw[<-] (3.3,1.5) to [out=95,in=250] (3.3,2.4) node [above] {{\scriptsize $Q_3$}};           
           
\draw [thick](0.2,-1.2) -- (1.4,-1.2) node[pos=.5,sloped,below] {$Y_2$} ;
\draw[thick]  (0.2,-1.3) -- (0.2,-1.1); 
\draw [thick](1.4,-1.2) -- (2.5,-1.2) node[pos=.5,sloped,below] {$T_2$} ;
\draw[thick]  (1.4,-1.3) -- (1.4,-1.1) 
                    (2.5,-1.3) -- (2.5,-1.1);
                    
\draw [thick](6,-1.2) -- (6.8,-1.2) node[pos=.5,sloped,below] {$Y_n$} ;
\draw[thick]  (6,-1.3) -- (6,-1.1); 
\draw [thick](6.8,-1.2) -- (7.4,-1.2) node[pos=.5,sloped,below] {$T_n$} ;
\draw[thick]  (6.8,-1.3) -- (6.8,-1.1) 
                    (7.4,-1.3) -- (7.4,-1.1);
 
\draw[thick] (0,0.5) -- (1,1.5) -- (1,0.8);
\draw[thick] (0.2,0) -- (2.5,2.3) -- (2.5,1.1) -- (4.4,3) -- (4.4,2.4) -- (4.7,2.7);
\draw[thick] (6.4,0.5) -- (7.4,1.5) -- (7.4,0.6);

\draw[thick,red] (1,0.8) -- (1,1.5);
\draw[thick,red] (2.5,1.1) -- (2.5,2.3);
\draw[thick,red] (4.4,2.4) -- (4.4,3);
\draw[thick,red] (7.4,0.6) -- (7.4,1.5);

\draw[dotted] (1,0) -- (1,0.8);
\draw[dotted] (2.5,0) -- (2.5,1.1);
\draw[dotted] (4.4,0) -- (4.4,2.4);
\draw[dotted] (7.4,0) -- (7.4,0.6);
\draw[dotted] (-0.5,0) -- (0,0.5);
\draw[dotted] (1.4,0) -- (2.5,1.1);
\draw[dotted] (2,0) -- (4.7,2.7);
\draw[dotted] (3.8,0) -- (5,1.2);
\draw[dotted] (5.9,0) -- (6.4,0.5);
\draw[dotted] (6.8,0) -- (7.4,0.6);

\draw[thick]  (5.2,-0.15) -- (5.2,0.15) 
                    (5.3,-0.15) -- (5.3,0.15);
\draw[white, fill=white!50] (5.21,-0.2) -- (5.21,0.2) -- (5.29,0.2) -- (5.29,-0.2) ;                   

\draw [decorate,decoration={brace,amplitude=5pt,mirror,raise=2pt},yshift=0pt] (1,0.8) -- (1,1.55) node [black,midway,yshift=0.3cm,xshift=0.35cm] {\scriptsize $D_1$};
\draw [decorate,decoration={brace,amplitude=5pt,mirror,raise=2pt},yshift=0pt] (2.47,1.1) -- (2.47,2.33) node [black,midway,yshift=0.1cm,xshift=0.39cm] {\scriptsize $D_2$};
\draw [decorate,decoration={brace,amplitude=4pt,mirror,raise=2pt},yshift=0pt] (4.36,2.4) -- (4.36,3) node [black,midway,yshift=0.1cm,xshift=0.45cm] {\scriptsize $D_3$};
\draw [decorate,decoration={brace,amplitude=5pt,mirror,raise=2pt},yshift=0pt] (7.4,0.6) -- (7.4,1.5) node [black,midway,xshift=0.5cm] {\scriptsize $D_n$};
\end{tikzpicture}}
		\label{fig:linear_value}
	}
	\subfloat[][]{
		 \scalebox{.65}{\begin{tikzpicture}[scale=0.90]
\draw[->] (0,0) -- (8.2,0) node[anchor=north] {time};
\draw[->] (0,0) -- (0,3.5) node[anchor=east] {$C(t)$};
\draw	(-0.3,0.25) node[anchor=south] {$C_0$};

 \draw[fill=gray!10] (1.4,0) to[bend right=20] (4.4,3) -- (4.4,2.1) to[bend left=17] (2,0);
\draw	(-0.5,0) node[anchor=north] {$t_0$}
           (0.2,0) node[anchor=north] {$t_1$}
		    (1.4,0) node[anchor=north] {$t_2$}
		    (2,0) node[anchor=north] {$t_3$}
		    (3.8,0) node[anchor=north] {$t_4$}
		    (6,0) node[anchor=north] {$t_{n-1}$}
		    (6.8,0) node[anchor=north] {$t_n$};
\draw[->,>=stealth]    (1,0) -- (1,-0.4) node[anchor=south,below] {$t'_1$};
\draw[->,>=stealth]  (2.5,0) -- (2.5,-0.4) node[anchor=south,below] {$t'_2$};
\draw[->,>=stealth]  (4.4,0) -- (4.4,-0.4) node[anchor=south,below] {$t'_3$};
\draw[->,>=stealth]   (7.4,0) -- (7.4,-0.4) node[anchor=south,below] {$t'_n$};
		    	    		 
\draw	(0.4,1.2) node{{\scriptsize $Q_1$}}
		    (2,0.8) node{{\scriptsize $Q_2$}};
\draw   (4.0,0.85) node{{\scriptsize $Q_4$}}
           (7,0.4) node{{\scriptsize $Q_n$}};
           
\draw[<-] (3.35,1.0) to [out=95,in=250] (3.35,1.9) node [above] {{\scriptsize $Q_3$}};         
           
\draw [thick](0.2,-1.2) -- (1.4,-1.2) node[pos=.5,sloped,below] {$Y_2$} ;
\draw[thick]  (0.2,-1.3) -- (0.2,-1.1); 
\draw [thick](1.4,-1.2) -- (2.5,-1.2) node[pos=.5,sloped,below] {$T_2$} ;
\draw[thick]  (1.4,-1.3) -- (1.4,-1.1) 
                    (2.5,-1.3) -- (2.5,-1.1);
                    
\draw [thick](6,-1.2) -- (6.8,-1.2) node[pos=.5,sloped,below] {$Y_n$} ;
\draw[thick]  (6,-1.3) -- (6,-1.1); 
\draw [thick](6.8,-1.2) -- (7.4,-1.2) node[pos=.5,sloped,below] {$T_n$} ;
\draw[thick]  (6.8,-1.3) -- (6.8,-1.1) 
                    (7.4,-1.3) -- (7.4,-1.1);
 
\draw[thick,red]  (1,1.5) -- (1,0.45);
\draw[thick,red] (2.5,2.3) -- (2.5,0.6);
\draw[thick,red] (4.4,3) -- (4.4,2.1);
\draw[thick,red] (7.4,1.5) -- (7.4,0.3);

\draw[dotted]  (-0.5,0) to[bend right=20] (1,1.5); 
\draw[thick]  (0,0.27) to[bend right=15] (1,1.5); 

\draw[dotted]  (0.2,0) to[bend right=20] (2.5,2.3); 
\draw[thick]  (1,0.45) to[bend right=15] (2.5,2.3); 

\draw[dotted]  (1.4,0) to[bend right=20] (4.4,3); 
\draw[thick]  (2.5,0.6) to[bend right=15] (4.4,3.05); 

\draw[dotted]  (2,0) to[bend right=20] (4.7,2.7); 
\draw[thick]  (4.4,2.1) to[bend right=3] (4.7,2.7); 

\draw[dotted]  (5.9,0) to[bend right=20] (7.4,1.5); 
\draw[thick]  (6.4,0.27) to[bend right=17] (7.4,1.5); 

\draw[dotted] (1,0) -- (1,0.5);
\draw[dotted] (2.5,0) -- (2.5,2);
\draw[dotted] (4.4,0) -- (4.4,2.4);
\draw[dotted] (7.4,0) -- (7.4,0.6);

\draw[dotted] (3.8,0) to[bend right=20] (5,1.2);
\draw[dotted] (6.8,0) to[bend right=20] (7.4,0.3);

\draw[thick]  (5.2,-0.15) -- (5.2,0.15) 
                    (5.3,-0.15) -- (5.3,0.15);
\draw[white, fill=white!50] (5.21,-0.2) -- (5.21,0.2) -- (5.29,0.2) -- (5.29,-0.2) ;                   

\draw [decorate,decoration={brace,amplitude=5pt,mirror,raise=2pt},yshift=0pt] (0.95,0.47) -- (0.95,1.5) node [black,midway,yshift=0.3cm,xshift=0.39cm] {\scriptsize $D_1$};
\draw [decorate,decoration={brace,amplitude=5pt,mirror,raise=2pt},yshift=0pt] (2.47,0.62) -- (2.47,2.3) node [black,midway,xshift=0.39cm] {\scriptsize $D_2$};
\draw [decorate,decoration={brace,amplitude=4pt,mirror,raise=2pt},yshift=0pt] (4.37,2.1) -- (4.37,3) node [black,midway,xshift=0.5cm] {\scriptsize $D_3$};
\draw [decorate,decoration={brace,amplitude=5pt,mirror,raise=2pt},yshift=0pt] (7.4,0.27) -- (7.4,1.5) node [black,midway,xshift=0.5cm] {\scriptsize $D_n$};              
\end{tikzpicture}}
		\label{fig:exponential_value}
	}
	\subfloat[][]{
	     \scalebox{.65}{	\begin{tikzpicture}[scale=0.90]
\draw[->] (0,0) -- (8.2,0) node[anchor=north] {time};
\draw[->] (0,0) -- (0,3.5) node[anchor=east] {$C(t)$};
\draw	(-0.3,0.25) node[anchor=south] {$C_0$};

 \draw[fill=gray!10] (1.4,0) to[bend left=20] (4.4,3) -- (4.4,2.55) to[bend right=19] (2,0);
\draw	(-0.5,0) node[anchor=north] {$t_0$}
           (0.2,0) node[anchor=north] {$t_1$}
		    (1.4,0) node[anchor=north] {$t_2$}
		    (2,0) node[anchor=north] {$t_3$}
		    (3.8,0) node[anchor=north] {$t_4$}
		    (6,0) node[anchor=north] {$t_{n-1}$}
		    (6.8,0) node[anchor=north] {$t_n$};
\draw[->,>=stealth]    (1,0) -- (1,-0.4) node[anchor=south,below] {$t'_1$};
\draw[->,>=stealth]  (2.5,0) -- (2.5,-0.4) node[anchor=south,below] {$t'_2$};
\draw[->,>=stealth]  (4.4,0) -- (4.4,-0.4) node[anchor=south,below] {$t'_3$};
\draw[->,>=stealth]   (7.4,0) -- (7.4,-0.4) node[anchor=south,below] {$t'_n$};
		    	    		 
\draw	(0.4,1.7) node{{\scriptsize $Q_1$}}
		    (1.4,0.9) node{{\scriptsize $Q_2$}};
\draw   (4.0,1.3) node{{\scriptsize $Q_4$}}
           (6.8,0.5) node{{\scriptsize $Q_n$}};
           
\draw[<-] (3.3,2.05) to [out=95,in=250] (3.3,2.95) node [above] {{\scriptsize $Q_3$}};            
           
\draw [thick](0.2,-1.2) -- (1.4,-1.2) node[pos=.5,sloped,below] {$Y_2$} ;
\draw[thick]  (0.2,-1.3) -- (0.2,-1.1); 
\draw [thick](1.4,-1.2) -- (2.5,-1.2) node[pos=.5,sloped,below] {$T_2$} ;
\draw[thick]  (1.4,-1.3) -- (1.4,-1.1) 
                    (2.5,-1.3) -- (2.5,-1.1);
                    
\draw [thick](6,-1.2) -- (6.8,-1.2) node[pos=.5,sloped,below] {$Y_n$} ;
\draw[thick]  (6,-1.3) -- (6,-1.1); 
\draw [thick](6.8,-1.2) -- (7.4,-1.2) node[pos=.5,sloped,below] {$T_n$} ;
\draw[thick]  (6.8,-1.3) -- (6.8,-1.1) 
                    (7.4,-1.3) -- (7.4,-1.1);
 
\draw[thick,red]  (1,1.48) -- (1,1.25);
\draw[thick,red] (2.5,2.3) -- (2.5,1.7);
\draw[thick,red] (4.4,3) -- (4.4,2.55);
\draw[thick,red] (7.4,1.5) -- (7.4,0.4);

\draw[dotted]  (-0.5,0) to[bend left=20] (1,1.5); 
\draw[thick]  (0,0.8) to[bend left=14] (1,1.48); 

\draw[dotted]  (0.2,0) to[bend left=20] (2.5,2.3); 
\draw[thick]  (1,1.25) to[bend left=11] (2.5,2.3); 

\draw[dotted]  (1.4,0) to[bend left=20] (4.4,3); 
\draw[thick]  (2.5,1.7) to[bend left=10] (4.4,3.0); 

\draw[dotted]  (2,0) to[bend left=20] (4.7,2.7); 
\draw[thick]  (4.4,2.55) to[bend left=3] (4.7,2.7); 

\draw[dotted]  (5.9,0) to[bend left=20] (7.4,1.5); 
\draw[thick]  (6.4,0.8) to[bend left=11] (7.4,1.5); 

\draw[dotted] (1,0) -- (1,1.25);
\draw[dotted] (2.5,0) -- (2.5,2);
\draw[dotted] (4.4,0) -- (4.4,2.4);
\draw[dotted] (7.4,0) -- (7.4,0.6);

\draw[dotted] (3.8,0) to[bend left=20] (5,1.2);
\draw[dotted] (6.8,0) to[bend left=20] (7.4,0.4);

\draw[thick]  (5.2,-0.15) -- (5.2,0.15) 
                    (5.3,-0.15) -- (5.3,0.15);
\draw[white, fill=white!50] (5.21,-0.2) -- (5.21,0.2) -- (5.29,0.2) -- (5.29,-0.2) ;                   

\draw [decorate,decoration={brace,amplitude=2pt,mirror,raise=2pt},yshift=0.5pt] (0.95,1.26) -- (0.95,1.5) node [black,midway,yshift=0.4cm,xshift=0.29cm] {\scriptsize $D_1$};
\draw [decorate,decoration={brace,amplitude=4pt,mirror,raise=2pt},yshift=0pt] (2.47,1.75) -- (2.47,2.3) node [black,midway,yshift=0.25cm,xshift=0.32cm] {\scriptsize $D_2$};
\draw [decorate,decoration={brace,amplitude=3pt,mirror,raise=2pt},yshift=0pt] (4.37,2.57) -- (4.37,3) node [black,midway,yshift=0.05cm,xshift=0.4cm] {\scriptsize $D_3$};
\draw [decorate,decoration={brace,amplitude=5pt,mirror,raise=2pt},yshift=0pt] (7.4,0.4) -- (7.4,1.5) node [black,midway,xshift=0.5cm] {\scriptsize $D_n$};              
\end{tikzpicture}}
		\label{fig:logarithmic_value}
	}
	\caption{Example of linear, exponential and logarithmic CoUD evolution.}
\end{figure*}

Finally, to explore a wide array of potential uses of the notion of \emph{cost}, we explore in this paper three sample cases for the $f_s(\cdot)$ function
\begin{align}
f_s(t) =
\begin{cases} 
\alpha t \\ 
e^{\alpha  t}  -1\\
\log(\alpha  t +1)
\end{cases} 
\end{align}
 for $\alpha \geq 0$.
As discussed earlier, we can not leverage CoUD if we do not assume that the samples of the observed stochastic process are correlated.
Thus, we propose the adjustment of CoUD according to the autocorrelation of the process. 
Specifically, if the autocorrelation is small we suggest the exponential function, while if the autocorrelation is large the logarithmic function is more suitable. 
For intermediate values the linear case is a reasonable choice.

The autocorrelation $R(t_1,t_2)=\mathbb{E}[x(t_1) x^*(t_2)]$ of a stochastic process is a positive definite function, that is $\sum_{i,j} \beta_i \beta_j^* R(t_i,t_j) >0$, for any $\beta_i$ and $\beta_j$.
Tuning the parameter $\alpha$ properly enables us to associate with accuracy the right $f_s(\cdot)$ function to a corresponding autocorrelation.
Next, we focus on VoIU and analyze it for each case of $f_s(t)$ separately. 
\subsection{Value of Information of Update Analysis}
We first derive useful results for the general case without considering specific queueing models.
For the first case, $f_s(t)=\alpha t$, expression (\ref{eq:V_i}) yields  
\begin{equation}
V_{P,i} = \frac{Y_i}{Y_i + T_i}.
\label{eq:Vi_at}
\end{equation}
Note that for $\alpha = 1$ the cost of update delay corresponds to the timeliness of each status update arriving and is the so called \emph{age of information}.
The cost reductions $\{ D_1, \dots , D_n \}$, depicted in Fig.~\ref{fig:linear_value}, correspond to the interarrival times $\{ Y_1, \dots , Y_n \}$, and also the limits
\begin{equation}
\lim_{Y_i \to +\infty} V_{P,i} = 1,
\label{eq:lim_Vi_at}
\end{equation}
\begin{equation}
\lim_{T_i \to +\infty} V_{P,i} = 0,
\label{eq:lim_Vi_at2}
\end{equation}
agree with the definition.
Next, for $f_s(t)=e^{\alpha  t} -1$, shown in Fig.~\ref{fig:exponential_value}, the definition of VoIU is 
\begin{equation}
V_{E,i} = \frac{e^{\alpha (Y_i + T_i)} - e^{\alpha T_i}}{e^{\alpha (Y_i + T_i)}-1},
\label{eq:Vi_expat}
\end{equation} 
and the corresponding limits are $\lim_{Y_i \to +\infty} V_{E,i} = 1$, $\lim_{T_i \to +\infty} V_{E,i} = 1-e^{-\alpha Y_i}$.

At last, for the case $f_s(t)=log(\alpha t +1)$, depicted in  Fig.~\ref{fig:logarithmic_value}, we obtain
\begin{equation}
V_{L,i} =\frac{\log(\alpha (Y_i + T_i) +1) - \log(\alpha T_i+1)}{\log(\alpha (Y_i + T_i) +1)}, 
\label{eq:Vi_logat}
\end{equation}
$\lim_{Y_i \to +\infty} V_{L,i} = 1$, and $\lim_{T_i \to +\infty} V_{L,i} = 0$.
The previous results can be interpreted as follows. 
As the interarrival times of the received packets become large the value of information of the updates takes its maximum value, underlining the importance to have a new update as soon as possible.
On the other hand, when the system time gets significantly large we expect that the received update is not as timely as we would prefer in order to maintain the freshness of the system, hence we assign to the VoIU metric the minimum value.

Suppose that our interval of observation is $(0,\mathcal{T})$.
Then, the time average value of information (normalized by the duration of time interval) is given by 
\begin{equation}
V_{\mathcal{T}} = \frac{1}{\mathcal{T}} \sum_{i=1}^{N(\mathcal{T})} V_i.
\label{eq:av_value}
\end{equation} 
Without loss of generality we assume that the first packet generation was at the time instant $t_0$ and the observation begins at $t = 0$ with an empty queue and the value $C(0) = C_0$.
Moreover, the observation interval ends with the service completion of $N(\mathcal{T})$ samples, with $N(\mathcal{T}) = max \{n \,|\, t_n \leq \mathcal{T}\}$ denoting the number of arrivals by time $\mathcal{T}$.

The time average value in \eqref{eq:av_value} is an important metric taken into consideration when evaluating the performance of a network of status updates and should be calculated for each case of $f_s(t)$ separately. 
The time average VoIU for the three considered cases can be rewritten as 
\begin{equation}
V_\mathcal{T} =  \frac{N(\mathcal{T})}{\mathcal{T}} \frac{1}{N(\mathcal{T})} \sum^{N(\mathcal{T})}_{i=1} V_i.
\label{eq:V_i2}
\end{equation}
Additionally, defining the effective arrival rate as 
\begin{equation}
\lambda = \lim_{\mathcal{T} \to \infty} \frac{N(\mathcal{T})}{\mathcal{T}} 
\label{eq:lambda}
\end{equation}
and noticing that $N(\mathcal{T}) \to \infty$ as $\mathcal{T} \to \infty$, and that the sample average will converge to its corresponding stochastic average due to the assumed ergodicity of $V_i$, we conclude with the 
expression 
 \begin{equation}
V = \lim_{\mathcal{T} \to \infty} V_\mathcal{T} = \lambda \:\mathbb{E} \big[ V \big],
\label{eq:av_V}
\end{equation}
where $\mathbb{E}[\cdot]$ is the expectation operator.
\section{Cost of Update Delay computation for the M/M/1 System}
\label{sec:avCoUD}
For a M/M/1 system status updates are generated according to a Poisson process with mean $\lambda$ and thus the interarrival times $Y_i$ are independent and identically distributed i.i.d exponential random variables with $\mathbb{E}[Y] = 1/\lambda$.
Furthermore, the service times are i.i.d. exponentials with mean $1/\mu$ and the server utilization is $\rho = \frac{\lambda}{\mu}$.

Additionally, the probability density function of the system time $T$ for the M/M/1 is \cite{Papoulis}
 \begin{equation}
\mathbb{P}_T(t) = \mu(1-\rho) e^{-\mu(1-\rho)t}, \:\:\: t \geq 0. 
\label{eq:system_time}
\end{equation}
Note that the variables $Y$ and $T$ are dependent and this complicates the calculations of the average cost of update delay in the general case.
The time average \emph{CoUD} of (\ref{eq:cost_t}) in this scenario can be calculated as the  sum of the disjoint $Q_1$, $Q_i$ for $i \geq 2$, and the area of width $T_n$ over the time interval  $(t_n, t'_n)$.
This decomposition yields 
\begin{equation}
C_\mathcal{T} = \frac{Q_1 + T^2_n/2 + \sum^{N(\mathcal{T})}_{i=2}Q_i}{\mathcal{T}}.
\label{eq:Delta_t2}
\end{equation}  
Below we derive the average \emph{CoUD} for the three cases of the $f_s(t)$ function that we have considered and find the optimum server policy for each one of them.

For $f_s(t)=\alpha t$, the area $Q_i$ for $i \geq 2$ is a trapezoid equal to the difference of two triangles, hence  
\begin{equation}
Q_{P,i} = \frac{1}{2}\alpha(T_i + Y_i)^2 - \frac{1}{2} \alpha T_i^2 = \alpha \big[Y_i T_i + \frac{Y_i^2}{2} \big].
\label{eq:Qi_at}
\end{equation}
Next, for $f_s(t)=e^{\alpha  t} -1$, the area  $Q_i$ yields
\begin{align}
Q_{E,i} &=  \int^{t'_i}_{t_{i-1}}  (e^{\alpha (t-t_{i-1})} -1) \,\mathrm{d}t -  \int_{t_i}^{t'_i}  (e^{\alpha  (t - t_i)} -1) \,\mathrm{d}t = \notag \\
&= \frac{1}{\alpha}  \big[ e^{\alpha (Y_i + T_i)} - e^{\alpha T_i} \big] - Y_i.
\label{eq:Qi_exp(t)}
\end{align}
And lastly, for $f_s(t)=log(\alpha t +1)$ we obtain
\begin{equation*}
Q_{L,i} =  \int^{t'_i}_{t_{i-1}}  log((\alpha (t-t_{i-1})+1) \,\mathrm{d}t -  \int_{t_i}^{t'_i}  log(\alpha (t-t_i) +1) \,\mathrm{d}t 
\label{eq:Qi_log}
\end{equation*}
\begin{equation}
\resizebox{.99\hsize}{!}{$ =\frac{1}{\alpha}  \big[ (\alpha (Y_i + T_i) +1) \log(\alpha (Y_i + T_i) +1) - (\alpha T_i+1) \log(\alpha T_i +1) \big] - Y_i. $}
\label{eq:Qi_log2}
\end{equation}
The time average \emph{CoUD} for the three cases can be rewritten as 
\begin{equation}
C_\mathcal{T} = \frac{\tilde{Q}}{\mathcal{T}} + \frac{N(\mathcal{T})-1}{\mathcal{T}} \frac{1}{N(\mathcal{T})-1} \sum^{N(\mathcal{T})}_{i=2} Q_i
\label{eq:Delta_3}
\end{equation}
where, $\tilde{Q} = Q_1 + T_n^2/2$ is a term that will vanish as $\mathcal{T} \to \infty$. 
Then, similarly to the VoIU analysis, we conclude with the expression 
 \begin{equation}
C = \lim_{\mathcal{T} \to \infty} C_\mathcal{T} = \lambda \mathbb{E} \big[ Q \big].
\label{eq:av_Delta}
\end{equation}
For the linear case, using the result of \cite{Kaul12_INFOCOM}, we obtain 
 \begin{equation}
C_{P} = \alpha \frac{1}{\mu} \left( 1+\frac{1}{\rho}+\frac{\rho^2}{1-\rho} \right) .
\label{eq:av_C_lin}
\end{equation}
Next, for the exponential case we compute the terms  
\begin{align}
 \mathbb{E}\big[e^{\alpha  T} \big] =\begin{cases} 
 \frac{-\mu(1-\rho)}{\alpha -\mu(1-\rho)}&\mbox{, if } \alpha -(\mu -\lambda) < 0 \\
 +\infty&\mbox{, otherwise},
\end{cases} 
\label{eq:expectation_Veat}
\end{align}
\begin{align}
\mathbb{E}\big[e^{\alpha (Y+T)} \big] =\begin{cases} 
 \frac{\mu(1-\rho)\lambda}{[\alpha -\mu(1-\rho)](\alpha-\lambda)}&\mbox{, if } \alpha -\lambda < 0 \\
 +\infty&\mbox{, otherwise}.
\end{cases} 
\label{eq:expectation_Veat2}
\end{align}
After applying all the relevant expressions to \eqref{eq:av_Delta}, we find the average CoUD to be   
 \begin{equation}
C_{E} = \frac{1}{\alpha} \lambda \left( \frac{\mu(1-\rho)}{\alpha-\mu(1-\rho)} \left( \frac{\lambda}{\alpha-\lambda}+1 \right) \right) -1.
\label{eq:av_C_exp}
\end{equation}
Finally, for the logarithmic case we compute some terms of equation \eqref{eq:Qi_log2} and others are evaluated numerically. We omit the details due to space limit.
\section{Value of Information Update Computation for the M/M/1 System}
\label{sec:avVoI}
Following the same procedure as in the CoUD metric we compute the average VoIU given by (\ref{eq:av_V}), for the M/M/1 queue with a first-come-first-served discipline.
For the $f_s(t)=\alpha t$ case, the expected value $\mathbb{E}[V]$ conditioned on the interarrival time $X=x$ can be obtained as
\begin{align}
& \mathbb{E}\left[ \frac{X}{X+T}\Big/ X=x \right] = -x\mu(1-\rho) e^{x\mu(1-\rho)} \notag \\
& \times Ei(-\mu(1-\rho)x)) \mbox{, for}\:\:\:(\mu-\lambda)>0.  \label{eq:expectation_Vat}
\end{align}
Further, using the iterated expectation and the probability density function of $X$, \eqref{eq:expectation_Vat} implies
\begin{align}
\mathbb{E}\big[V_P\big] = \frac{\mu(1-\rho)}{2\lambda} \: {}_2 F_1(1,2{;}3{;}\frac{2 \lambda -\mu}{\lambda}),
\label{eq:av_V_lin}
\end{align}
where the integral is calculated with the help of \cite[6.228]{table} and ${}_2 F_1$ is the hypergeometric function \cite{wolfram}.
\begin{figure*}[ht]
	\centering
	\subfloat[][]{
		\includegraphics[draft=false,scale=.34]{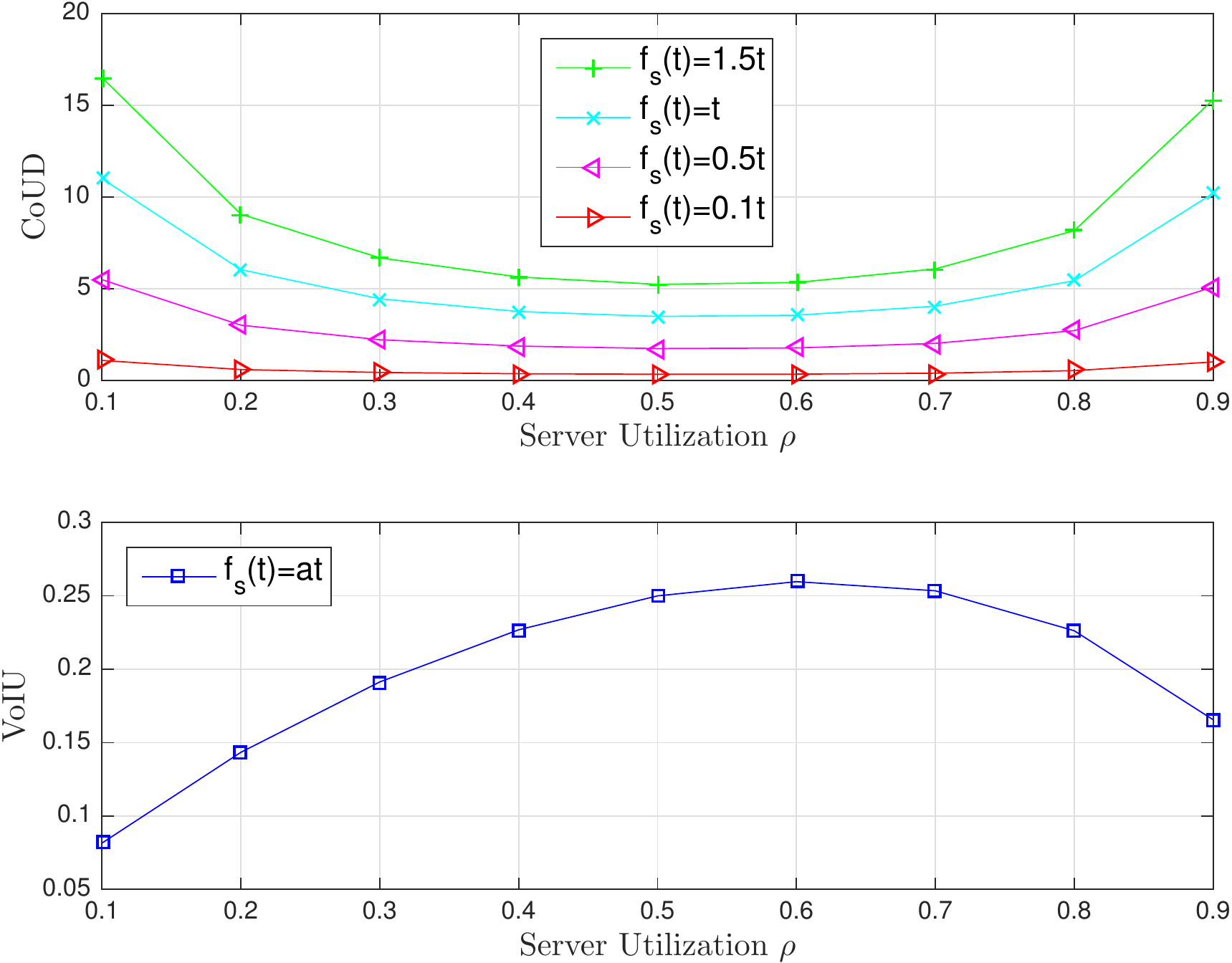}
		\label{fig:CoUDandVoI_vs_utilization_linear}
	}
	\subfloat[][]{
		\includegraphics[draft=false,scale=.34]{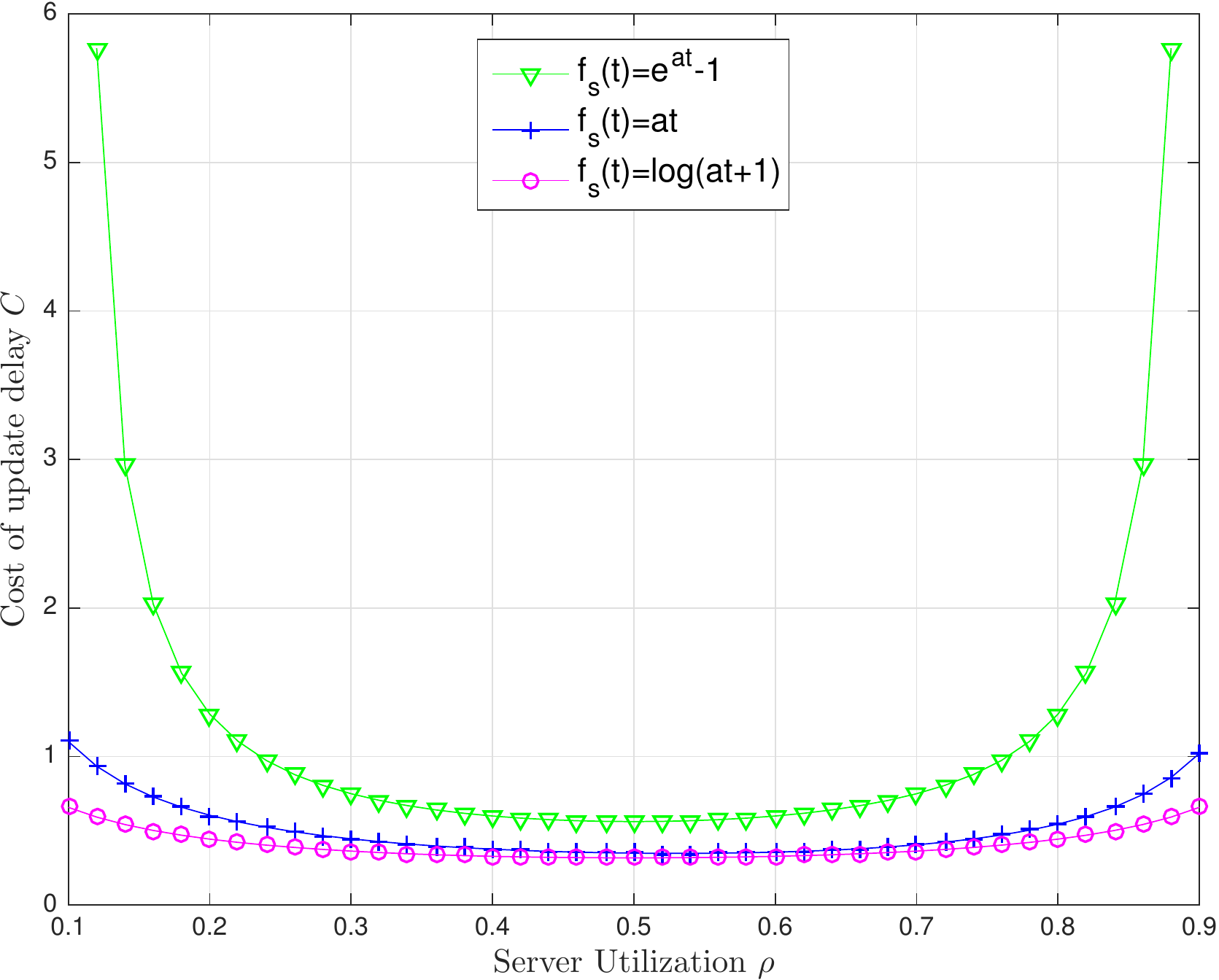}
		\label{fig:CoUD_vs_utilization}
	}
	\subfloat[][]{
		\includegraphics[draft=false,scale=.34]{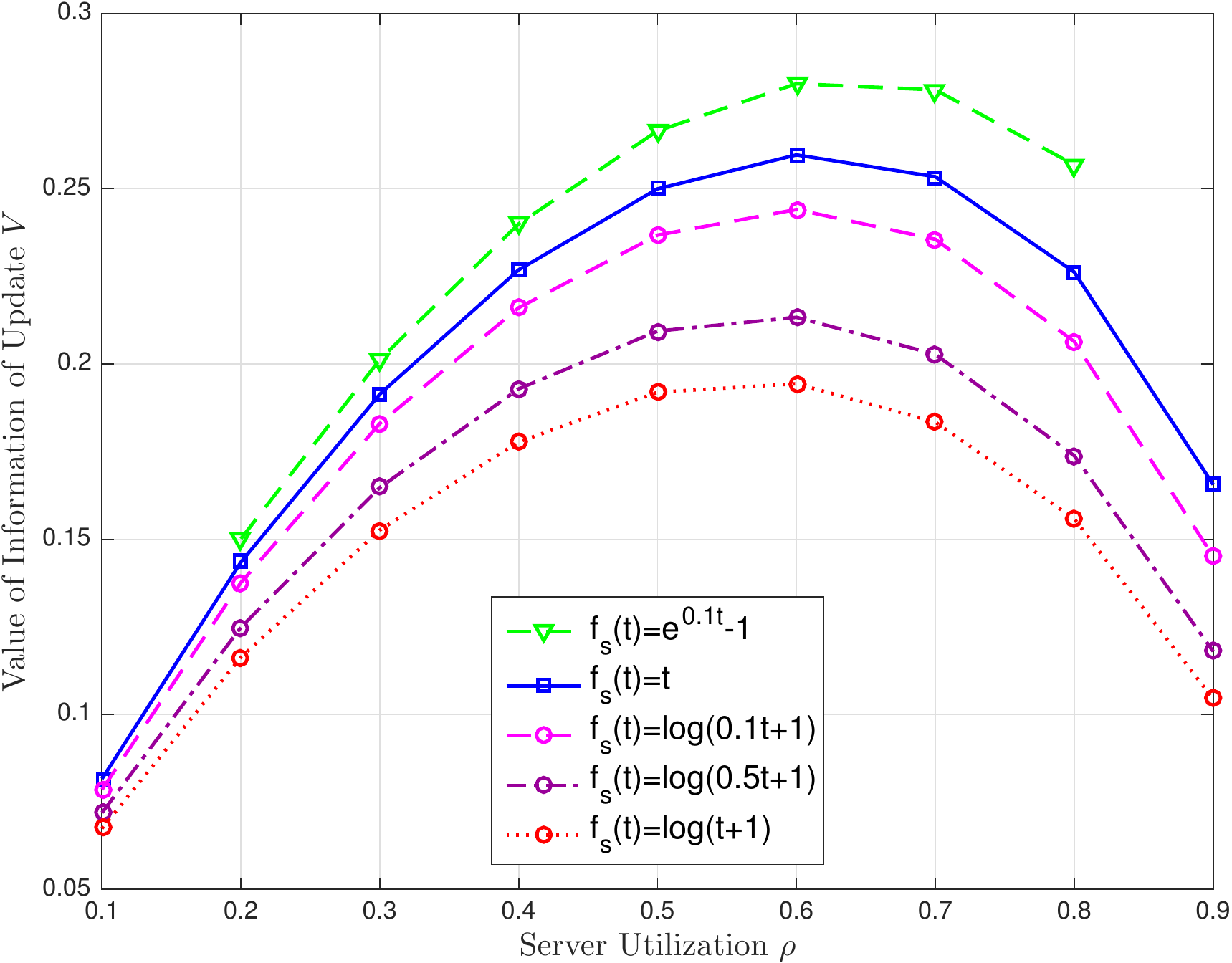}
		\label{fig:VoI_vs_utilization}
	}
	\caption{Comparison of the average CoUD and VoIU vs. utilization for the M/M/1 system with $\mu=1$, and different function cases.}
\end{figure*}

Next, for the cases $f_s(t)=e^{\alpha  t} -1$ and $f_s(t)=\log(\alpha t+1)$, we compute numerically the expected values $\mathbb{E}\big[V_E\big]$ and $\mathbb{E}\big[V_L\big]$.
\section{Numerical Results}
\label{sec:results}
In this section, we evaluate the performance in terms of the CoUD and VoIU metrics, as calculated in the previous sections.
We consider a M/M/1 system model with service rate $\mu=1$, therefore the server utilization $\rho$ is equal to the arrival rate $\lambda$.

In Figure~\ref{fig:CoUDandVoI_vs_utilization_linear}, we illustrate the variation of the average CoUD and VoIU with the server utilization $\rho$, for the linear case.
Recall that for $f_s(t)=\alpha  t$ the VoIU is independent to the parameter $\alpha$, therefore for multiple CoUD curves corresponds only one VoIU curve.
This indicates that as far as the cost per time unit is linearly increased, higher cost leads to higher average CoUD, but the same average VoIU.
This is due to the fact that we assign to each unit of time the same cost.
Increasing $\alpha$ results to proportional increase of the average CoUD, however the optimal server policy is the same for every function.
Moreover, the optimal policy with respect to VoIU is different than the one for CoUD with the former being actually greater. 
Thus, it is more appropriate to increase rather than decrease $\rho$ in case the optimal utilization can not be achieved.

In Figure~\ref{fig:CoUD_vs_utilization} the average CoUD is depicted as a function of the server utilization for the linear, exponential, and logarithmic functions with parameter $\alpha=0.1$.
All three functions have similar behaviour, with the minimum CoUD achieved when $\rho \approx 0.5$.
Over all values of $\rho$, the exponential $f_s$ has the greatest CoUD, followed by the linear $f_s$ and then the logarithmic $f_s$, that is, $C_E>C_P>C_L$. 
However, as $\rho$ deviates from the optimum, we see that the exponential function becomes sharper that the linear function, and the logarithmic function becomes smoother.
For smaller utilizations where status updates are not frequent enough and for higher utilization where packets spend more time in the system due to backlogs, CoUD is increased.
For this increase each function sets its own cost per time unit that results to more intense growth for the exponential average CoUD and less intense growth for the logarithmic CoUD.

Figure~\ref{fig:VoI_vs_utilization} presents the numerical evaluation of the quantities $\lambda\mathbb{E}\big[V_P\big]$, $\lambda\mathbb{E}\big[V_E\big]$, and $\lambda\mathbb{E}\big[V_L\big]$ for three values of the parameter $\alpha$, $0.1$, $0.5$, and $1$.
As we shift from $f_s(t)= \log{t+1}$ to $f_s(t)=e^{0.1t}-1$ VoIU becomes greater over all $\rho$ and all functions follow similar behaviour.
For all cases, the maximum VoIU is achieved when $\rho \approx 0.6$.
Note that VoI is directly related to CoUD.
On the average analysis, taking the linear function as a point of reference, we see that choosing an exponential function would result in higher CoUD and VoIU, while choosing the logarithmic function would result in lower CoUD and VoIU.
This tradeoff considers timeliness against timeliness that is combined with transmission resources (i.e., bandwidth).
\section{Conclusion}
\label{sec:conclusions}
In this study, we have considered the characterization of the information transmitted over a source-destination network, modelled as a M/M/1 queue.
To capture freshness, we introduce the CoUD metric through three cost functions that can be chosen in relation with the autocorrelation of the process under observation.
To characterize the importance of an update, we define VoIU that measures the reduction of CoUD and therefore of uncertainty.
CoUD and VoIU can be used interchangeably depending on the application.
We  analysed the relation between CoUD and VoIU and observed that convex and concave CoUD functions lead to a tradeoff between CoUD and VoIU, while linearity reflects only on the CoUD.

Depending on the application we can choose the utilization that satisfies the minimum CoUD objective or the maximum VoIU objective. 
In the linear CoUD case, VoIU is independent to the cost assigned per time unit.
In the exponential and logarithmic cases however, there is a tradeoff between CoUD and VoIU.
That is, the smaller the average CoUD, the smaller the average VoIU. 
For high correlation among the samples, choosing $f_s(t)= \log{\alpha t+1}$ decreases their value of information and equivalently choosing $f_s(t)=e^{\alpha t}-1$ in low correlation has the opposite effect.



\bibliography{references}
\bibliographystyle{IEEEtran}

\end{document}